# Inventions on reducing number of keys on a Computer Keyboard
## -A TRIZ based analysis

**Umakant Mishra**

Bangalore, India

umakant@trizsite.tk

http://umakant.trizsite.tk

## Contents

## 1. Introduction

A computer keyboard consists of several sections and each section consists of several numbers of keys. The text entry section contains the standard character keys, navigation section contains cursor movement and page control keys, numeric keypad contains numeric keys and function keys section contain function keys and special keys. Although the increased number of keys helps smooth interaction with a computer there are situations where it is necessary to reduce the number of keys.

### 1.1 Need for reducing number of keys

- In portable and handheld computers there is not enough space to keep all the conventional keys. It is necessary to reduce the number of keys on such keyboards.

- The new users and children find the keyboard to be clumsy and difficult with so many numbers of keys.

- While we are using software like games we don't need most of the keys available on a standard keyboard.

- If the data is only numeric (or confined to only few characters), then it will be faster to enter data through a reduced number of keys than a full keyboard.

### 1.2 Problems in reducing the number of keys

- Reducing the number of keys does not allow us to generate all the required characters and functions by using separate keys. In some cases we may have to use two or more strokes to generate a character.

- Reducing the number of keys may need to change the standard key layout which will need relearning the keyboard operation for the existing users.

- A reduced number of keys may reduce the size of the keyboard which is not convenient to operate for a desktop computer.

### 1.3 Using TRIZ to reduce keys in keyboard

**Problem:** Generally more number of keys support more number of functions and less number of keys support less number of functions. We need more number of functions (all the functions available in a full size keyboard), but we want only less number of keys **(Contradiction).**

**Possible Solution:** Use multi-stroke per character to generate more number of command functions from less number of keys **(Principle-17: Another dimension)**.

**Problem:** In a multi-stroke mechanism the user has to press two or more keys to get a character, where the user is confused about which key to press after what. We want to use multi-stroke mechanism to generate more characters, but we don't want users to be confused **(Contradiction)**.

**Solution:** Provide an operating guide (or template) to assist users on selecting keystrokes **(Principle-8: Counterweight)**.

### 1.4 Some past inventions on reduced key keyboards

There are some reduced-key keyboards which try to achieve the full functionality of a regular keyboard by pressing two or more times for each character. Some inventions using such multi-stroke technologies are given below.

- U.S. Pat. No. 3833765 issued to E. Hillborn et al. Sept., 1974 provides for both alphabet characters and more complex messages to be chosen by the two letter codes provided by the two successive keystrokes from a set of twelve keys arranged for four finger touch typing with one hand.

- British Patent No. 1417849, Dec 1975, with two strokes per character, operation of a typewriter is made possible by means of only seven keys.

- U.S. Pat. No. 3892958, issued July 1975 to C. C. Tung, discloses a similar method wherein two prefix or shift keys change the keyboard role to provide three times as many input possibilities from a set of keys. The keyboard having 36 keys can access about 100 characters and functions.

- U.S. Pat. No. 3967273, issued to K. Knowlton, June 1976, provides for two keystroke per character entry operation from a telephone keyboard to generate all alphanumeric characters from twelve keys.

- U.S. Pat. No. 4,007,443 issued Feb. 8, 1977 to M. A. Bromberg et al. uses a set of twenty keys in four different modes selected by keys depressed by fingers on a hand holding keyboard to produce alphanumeric input data.

## 2. Inventions on reduced key Keyboards

### 2.1 Comprehensive computer data and control entries from very few keys operable in a fast touch typing mode (Patent 5062070)

**Background problem**

With the evolution of computers the processing power of the computers has gone up, but the speed of manual inputting of data through a keyboard has not improved because of various limitations of the keyboard.

The conventional QWERTY keyboards have several keys for each character but the efficiency is reduced because of inconvenient positions for finger movement. There are some methods to improve the input speed by reducing the number of keys, but the method has a problem of visual key selections.

It is necessary to reduce the number of keys, number of input strokes, typing time and convenience.

**Solution provided by Patent 5062070**

Lapeyre invented a keyboard (issued October 1991, assignee The Laitram Corporation) which has as few as only three or four keys operable by the fingers of a single hand. The keyboard can generate characters as many as a conventional keyboard.

The invention carries visual instructions for the keystroke sequence for each entry which reduces the difficulty of the operator. When the first keystroke is pressed the instructions display the choices available from the next keystrokes.

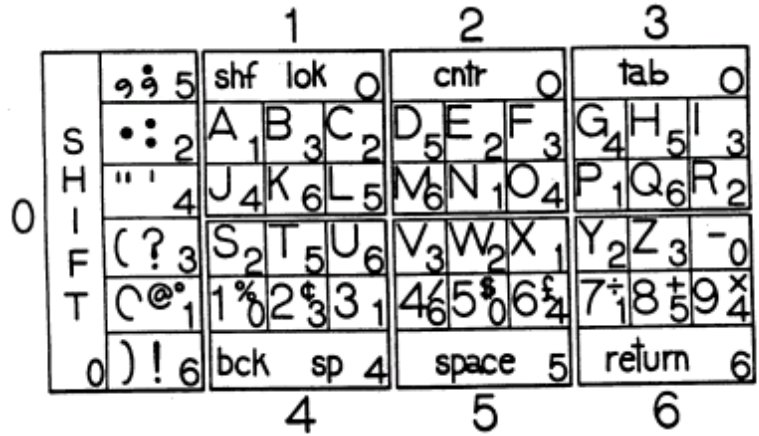

The invention provides high speed typing as standard QWERTY keyboard even by using only four keys.

**TRIZ based analysis**

Generally more number of keys supports more number of functions and less number of keys support less number of functions. We want only less number of keys, but we need more number of functions **(Contradiction)**.

The invention reduces the number of keys but get all the functions by increasing the strokes per character **(Principle-17: Another dimension)**.

The invention carries visual instructions for the keystroke sequence for each entry, which reduces the difficulty of the operator **(Principle-8: Counterweight)**.

The invention uses only four keys and allows all the fingers to always remain on their home position. **(Principle-39: Calm)**.

**2.2 Comprehensive computer data and control entries from very few keys operable in a fast touch typing mode (Patent 5184315)**

**Background problem**

There are so many possible functions performed by modern computers, which are not available to a keyboard which has less number of keys. If we use a two-stroke per entry keyboard 12 keys can produce as much as 144 characters.

There are devices that use small keypads to be operated with only one hand. This keyboard can contain only very few keys but should be capable of inputting large number of entries. It is necessary to improve such keyboards which may have less number of keys, but should support large number of functions.

**Solution provided by the invention**

Lapeyre disclosed a computer-keyboard system (patent 5184315, assignee Laitram corporation, issued Feb 1993) which has as few as three or four keys to be operated by a single hand but supports all input signals to replace a conventional typewriter keyboard.

The invention discloses a four key alphanumeric keyboard entry system providing forty-eight entries. Each key has a graphic key chart on the key showing symbolically twelve entry command choices available for each key. The solution provides entry of 26 alphabetic keys 10 numeric and other important keys with the help of only four keys.

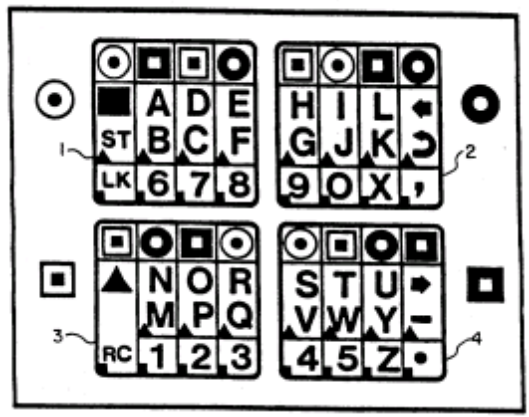

The invention is a unique type of key arrangement and displaying key guide, which can be applied on keyboards having only three, four, seven or nine keys. There may be two or mode modes to change the command assignment of the keys.

The key character layout is important in this invention which claims to increase the typing speed. The keys are particularly arranged for efficient tough typing input by allocation of the most frequently used alphabet characters to the easiest to stroke key pair locations.

**TRIZ based analysis**

The invention follows multi-stroke mechanism to generate more characters with less number of keys **(Principle-17: Another dimension)**.

Displaying key-guide on the top of the keys and using visual symbols for easy recognition **(Principle-8: Counterweight, Principle-32: Color change)**

## 2.3 Computer keyboard adapter providing large size key surfaces (Patent 5514855)

**Background problem**

There are certain educational programs available for young children which do not use most of the keys in a 101 key conventional keyboard. The large number of additional keys on the keyboard creates confusion and leads to incorrect response. There is a need to have a special keyboard for educational software for young children.

**Solution provided by the invention**

Sullivan developed an adapter for the keyboard (Patent 5514855, assignee Alpha Logic Inc, Issued May 1996), which amplifies specific keys of the keyboard to facilitate use by young children. The new keyboard has small number of large keys which when depressed, will cause depression of one or more keys of a selected area. There will be a software application which will associate the depressed key in the existing keyboard to determine the depressed key o the new keyboard.

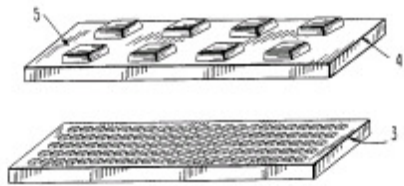

This new keyboard is intended to be placed on top of the existing keyboard so that other keys in the keyboard are protected from being depressed.

**TRIZ based analysis**

One solution is to protect all the keys on the keyboard with a thin hard cover except the specific keys which are required for the educational software **(Principle-2: Taking out)**. Another method is to change the labels or color of the specific key which are available to be used with the software **(Principle-32: Color change)**. However, these solutions do not reposition the required keys in an organized way for convenient access.

This invention provides an expanded interface for specific required keys **(Principle-37: Expansion)**.

The invention keeps the new keyboard on top of the existing keyboard **(Principle-7: Nested doll)**.

The invention uses a software to determine the pressed switch in the new keyboard from the pressed keys of the existing keyboard. **(Principle-36: Conversion)**.

## 2.4 A totally new approach for high speed keyboard (Patent 5828323)

**Background problem**

The standard computer keyboards typically contain 101 keys for inputting character sets including alphabets, numbers, symbols, or functions. These keyboards require coordinated movements of the user's arms, hand, and fingers, which limit the speed of operation and tire the user. It is necessary to simplify the keyboard to simplify typing and facilitate high speed typing with less manual effort?

**Solution provided by the invention**

Bartent invented a radically different keyboard (US Patent 5828323, Issued in Oct 98) which has only ten keys arranged in a curve to fit the ten corresponding fingers of a user. The invention was claimed to be ergonomic and high speed than standard computer keyboards.

The keys are activated in pairs to produce coding signals of a conventional keyboard. In other words, the user produces unique letters and functions by pressing two keys together.

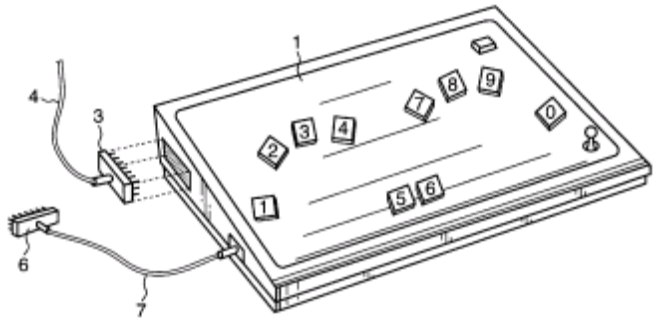

As the number of unique binary combinations of ten elements is 45, the ten-key keyboard system is capable of representing at least 26 characters plus the function keys of a standard 101-key keyboard. The invention provides a chart of key layout to be followed.

**TRIZ based analysis**

The keyboard should contain no keys (or minimum number of keys) and should need no (or minimum) movement of fingers and arm **(Ideal Final Result)**.

This invention reduces the number of keys to only 10 **(Principle-2: Taking out)**.

It organizes the keys in curved shape to align with the position of fingertips **(Principle-14: Curvature)**.

## 3. Summary and conclusion

There are different purposes to reduce the number of keys in a keyboard. Some of them intend to reduce the size of the keyboard while some others intend to increase the speed of typing.

Each of these invention use multi-stroke mechanism to generate more number of signals from less number of keys. The sequence of pressing keys in a multi-stroke mode is different for different inventions.

Reducing the number of keys is very useful in specific circumstances (e.g., in a mobile handset) especially where there is no space for keeping more keys.

## 4. Reference

1. US Patent 5062070, "Comprehensive computer data and control entries from very few keys operable in a fast touch typing mode", Invented by Lapeyre, assignee The Laitram Corporation, issued Oct 1991

2. US Patent 5184315, "Comprehensive computer data and control entries from very few keys operable in a fast touch typing mode", invented by Lapeyre, assigned by The Laitram Corporation, issued Feb 1993.

3. US Patent 5514855,"Computer keyboard adapter providing large size key surfaces", invented by Sullivan, assignee- Alpha Logic, Incorporated, issued may 1996

4. US Patent 5828323, "A totally new approach to high speed keyboard", Bartent, Oct 98

5. US Patent and Trademark Office (USPTO) site, http://www.uspto.gov/